\documentclass{iaus}
\usepackage{graphicx}
\def\mum{${\rm \mu m}$}
\newcommand{\HII}{H\,\textsc{ii}\,\,}

\title[The PAH bands.] 
{The PAH hypothesis after 25 years.}

\author[Els Peeters]   
{Els Peeters$^{1,2}$}

\affiliation{$^1$Department of Physics and Astronomy, University of Western Ontario \\ London, ON N6A 3K7, Canada\\ email: {\tt epeeters@uwo.ca} \\[\affilskip]
$^2$SETI Institute, 189 Bernardo Ave., \\ Suite 100, Mountain View, CA 94043, USA \\ email: {\tt epeeters@seti.org}}

\pubyear{2011}
\volume{280}  
\pagerange{}
\jname{The Molecular Universe}
\editors{}
\begin{document}

\maketitle

\begin{abstract}
The infrared spectra of many galactic and extragalactic objects are dominated by emission features at 3.3, 6.2, 7.7, 8.6 and 11.2 $\mu$m. The carriers of these features remained a mystery for almost a decade, hence the bands were dubbed the unidentified infrared (UIR) bands. Since the mid-80's, the UIR bands are generally attributed to the IR fluorescence of Polycyclic Aromatic Hydrocarbon molecules (PAHs) upon absorption of UV photons -- the PAH hypothesis. Here we review the progress made over the past 25 years in understanding the UIR bands and their carriers.
\end{abstract}

\firstsection 
\section{Introduction}

Soon after infrared (IR) detectors became available and opened up the infrared universe,  broad, resolved emission features were observed centered at 8.6 and 11.2 \mum\, (Gillet \etal\, 1973) and subsequently at 3.3, 6.2 and 7.7 \mum\, (Russel \etal\, 1978). These emission bands always appeared together and were found in planetary nebulae, \HII regions, and reflection nebulae. Since the carrier of these bands remained unidentified for almost a decade, they were dubbed the Unidentified InfraRed (UIR) bands. 

Duley \& Williams (1981) recognized that the UIR bands peak at wavelengths corresponding to the vibrational modes characteristic of aromatic material. This led to several proposed carriers of the UIR bands, including Hydrogenated Amorphous Carbon (HAC, e.g. Duley \& Williams 1983; Borghesi \etal\, 1987), Quenched Carbon Composites (QCC, e.g. Sakata \etal\, 1984), Polycyclic Aromatic Hydrocarbons (PAHs, e.g. Puget \& L\'eger 1984; Allamandola \etal\, 1985), coal (e.g. Papoular \etal\, 1989), nanodiamonds (Jones \& d'Hendecourt 2000), Rydberg matter (Holmlid 2000) and Locally Aromatic Polycyclic Hydrocarbons (Petrie \etal\, 2003). 
The main arguments against several proposed carriers are (i) the UIR bands are observed in reflection nebulae, and (ii) their color temperature is independent of the distance from the illuminating star (Sellgren 1984). Indeed, in these environments, classical dust grains are too cold to emit at mid-infrared wavelengths. Instead, these observations suggest that the carrier of the UIR bands attains very high temperature upon absorption of a single UV photon and hence small species, which have a limited heat capacity, are required (Sellgren 1984).

PAH molecules fulfill this requirement. Upon absorption of a single UV photon, PAHs reach a temperature of $\sim$1000K. They subsequently relax mainly by IR emission in their vibrational modes leading to the UIR bands. Specifically, the 3.3 \mum\, band is due to the CH stretching mode, the 6.2 \mum\, band to the CC stretching mode, the 7.7 \mum\, band to coupled CC stretching and CH in-plane bending modes, the 8.6 \mum\, band to the CH in-plane bending mode and the 10-15 \mum\, region to the CH out-of-plane (CH$_{oop}$) bending modes. PAHs contain a large fraction of the available carbon (10-20\%) and play an important role in various physical and chemical processes. Thus, the PAH hypothesis was born (Puget \& L\'eger 1984, 1989; Allamandola \etal\, 1985, 1989).

In this paper, we focus on the current status of the observational characteristics of the PAH emission bands\footnote{For the remaining of this paper, we refer to the UIR bands as the PAH bands.}, their dependence on the local physical conditions and the implications for the physical and chemical characteristics of the carriers. We also highlight the current and future use of these emission bands as a diagnostic tool.  

There is an extensive body of literature on PAHs and related species in astrophysical environments; we refer in particular to the reviews by Allamandola \etal\, (1989), Puget \& L\'eger (1989), Tielens (2005, 2008), and the proceedings: "PAHs in the Universe" (Joblin \& Tielens 2011).

\section{The rich PAH spectrum}

\begin{figure}[t]
\begin{center}
\includegraphics[width=12cm, angle=90]{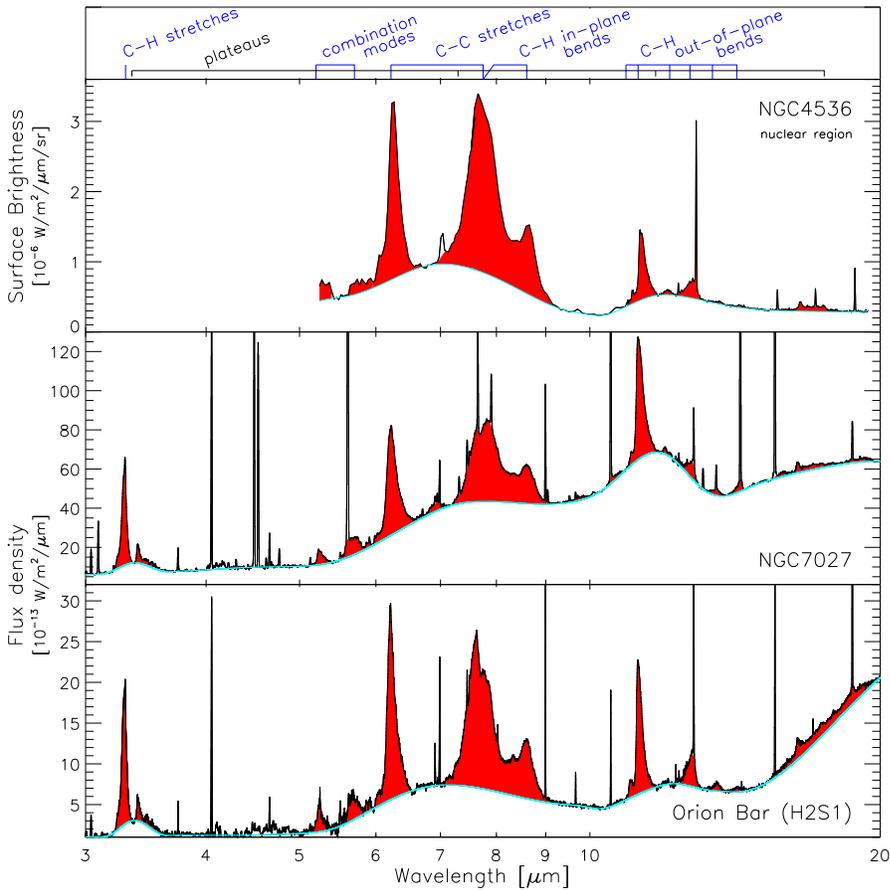}
 \caption{The ISO-SWS spectra of the planetary nebula NGC 7027 and the Photo-Dissociation Region (PDR) the Orion Bar, and the Spitzer-IRS spectrum of the nuclear region of NGC 4536, an HII-type galaxy, (SINGS legacy program; Kennicutt \etal\, 2003) illustrate the richness and variety of the PAH spectrum. Also indicated are the aromatic mode identifications of the major PAH bands. Figure adapted from Peeters 2011.}
   \label{peeters_spectrum}
\end{center}
\end{figure}

The Infrared Space Observatory (ISO) and the Spitzer Space Telescope showcased the richness and complexity of the astronomical PAH spectra. Beside the main bands at 3.3, 6.2, 7.7, 8.6, 11.2, and 12.7 \mum, a plethora of weaker bands is observed at 3.4, 3.5, 5.25, 5.75, 6.0, 6.6, 6.9, 7.2-7.4, 8.2, 10.5, 10.8, 11.0, 12.0, 13.5, 14.2, 15.8, 16.4, 16.6, 17.0, 17.2, 17.4, and 17.8 \mum\, (see Fig. \ref{peeters_spectrum}; \footnote{A band at 18.9 \mum\, has been attributed to PAH emission as well. However, this band is now assigned to the fullerene C$_{60}$ (Cami \etal\, 2010, Sellgren \etal\, 2010).}). 
In addition, deuterated PAHs, PADs, are tentatively detected at 4.4 and 4.65 \mum\, (Peeters \etal\, 2004a). 

The PAH bands are often found on top of broad plateaus located roughly at 3.2--3.6, 6--9, 11--14 and 15--20 \mum. These plateaus are part of the PAH spectrum. However, within the literature, they are treated in two different ways. One method considers them as part of the individual PAH emission bands, which are fitted by Lorentzian or Drude profiles (e.g. Boulanger \etal\, 1996; Smith \etal\, 2007; Galliano \etal\, 2008b). Alternatively, the plateaus are treated as features that are separate and distinct from the individual PAH bands. A local spline continuum is then determined to distinguish between both components (e.g. Hony \etal\, 2001, Peeters \etal\, 2002, van Diedenhoven \etal\, 2004). Neither approach is however technically correct (for a detailed discussion, see Tielens 2008). The applied method clearly determines the derived PAH band strengths and profiles. But fortunately, the overall conclusions are independent of the chosen approach (Uchida \etal\, 2000, Galliano \etal\, 2008b).

In addition to the complexity of the PAH spectrum, ISO and Spitzer revealed that the PAH bands and their carriers pervade the universe. They are observed in a large variety of astronomical environments including the (diffuse) ISM, reflection nebulae, T-Tauri stars, Herbig AeBe stars, \HII regions, post-AGB stars, planetary nebulae and, more exceptionally in few AGB stars, WR stars, novae, and supernovae remnants.  Moreover, they are widespread in various types of galaxies up to redshifts of $\sim$3 (spectroscopically) and $\sim$6 (photometrically). The PAH bands generally dominate the mid-IR spectra of these sources; a remarkable 10-15\% of the total power output of galaxies is emitted in these PAH bands (Smith \etal\, 2007). Clearly, the carriers of these infrared emission bands are abundantly present throughout the universe.

\section{Band Intensities}

\subsection{Observational facts}
\label{peeters_intensities}

\begin{figure}
\includegraphics[height=5.4cm]{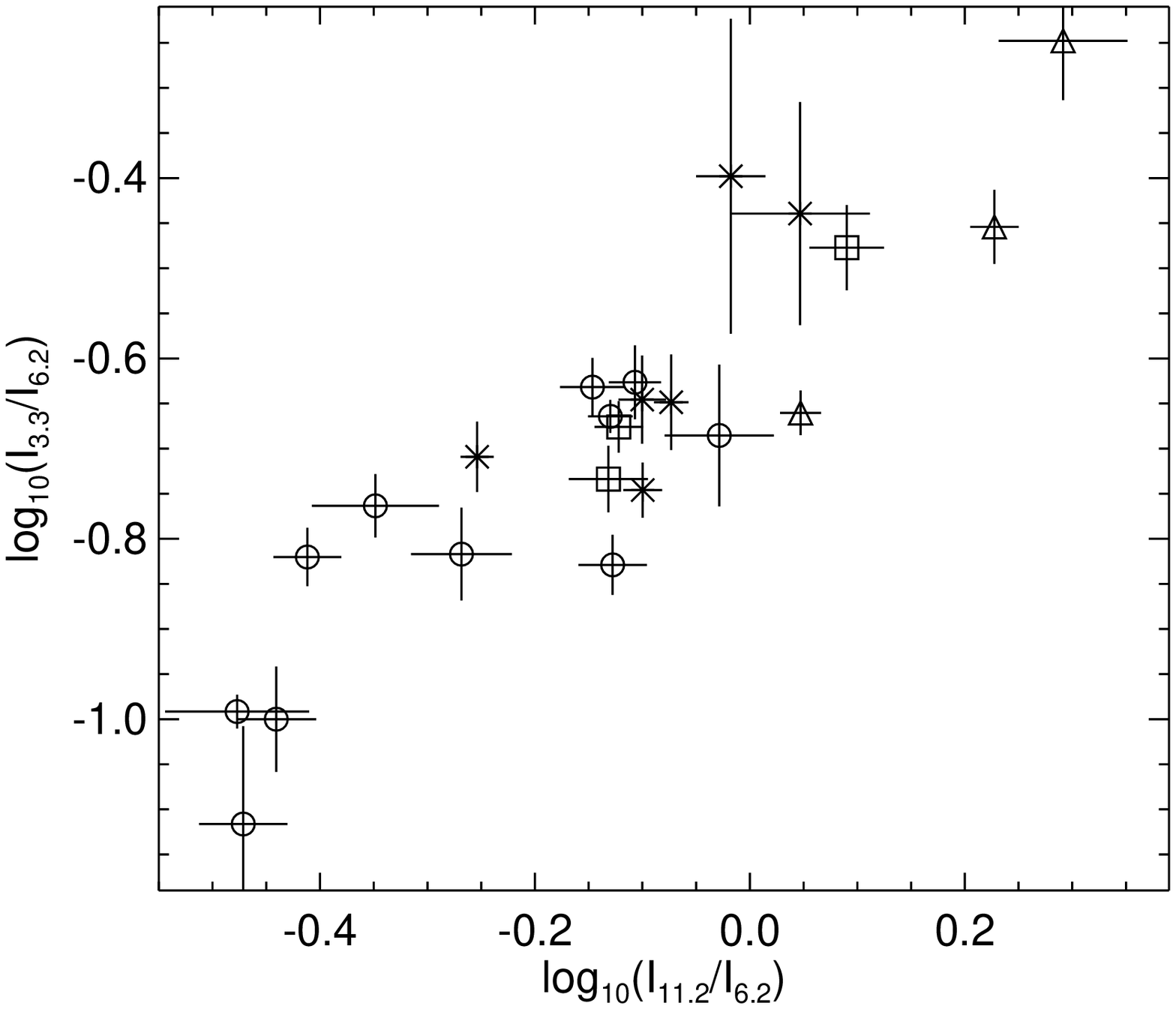}
\qquad
\includegraphics[height=5.9cm]{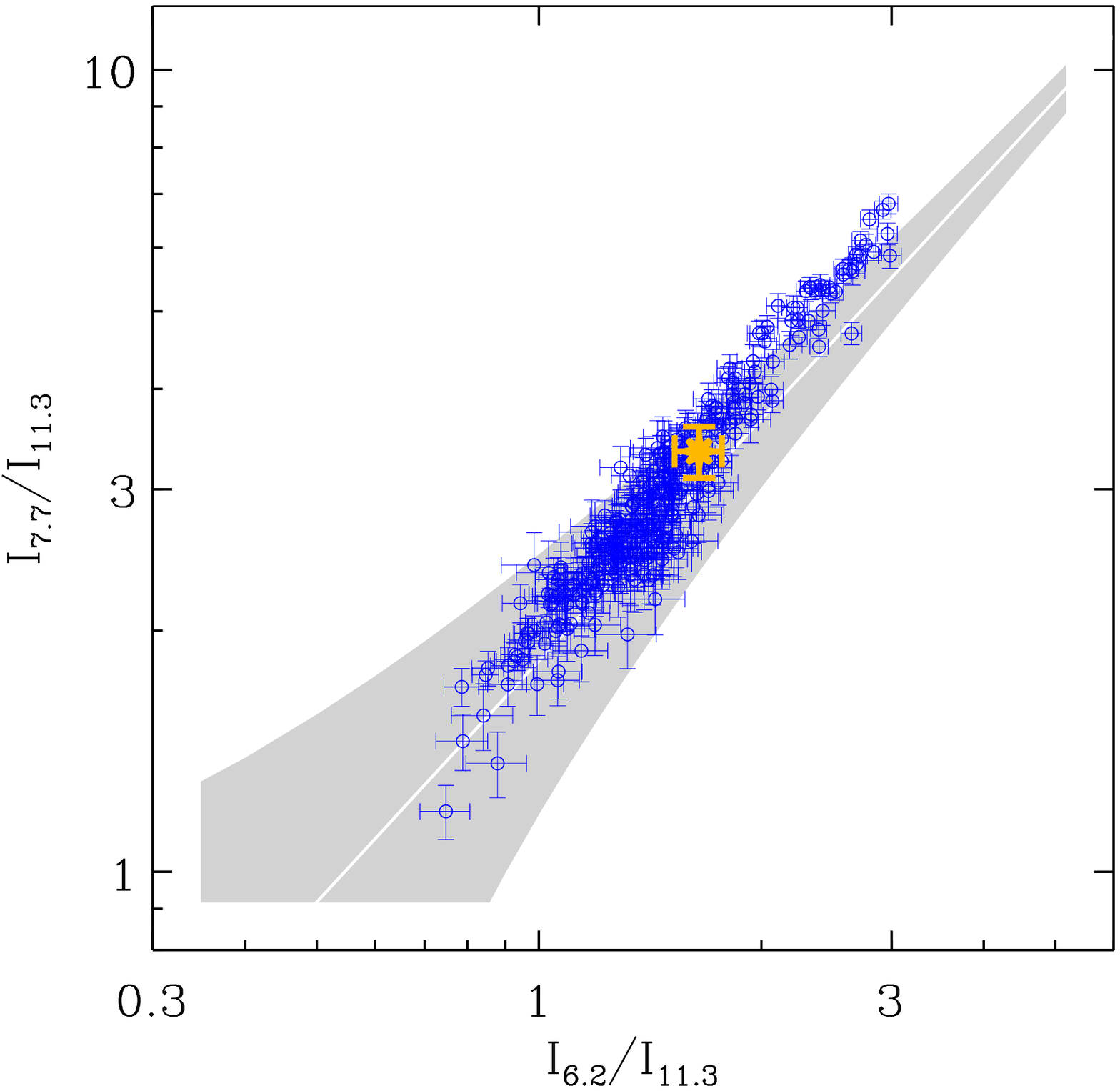}
\caption{{\bf Left:} The correlation of the 3.3 and 11.2 \mum\, band strengths normalized to the 6.2 \mum\, band strength. Hexagons represents \HII regions, stars intermediate mass star-forming regions, squares reflection nebulae and triangles are planetary nebulae. Figure from Hony \etal\, 2001. {\bf Right:} The variation in the CC/CH band ratios across the starburst galaxy M82. The grey filled area represent the correlation obtained for the integrated spectra of a large sample of galactic and extragalactic sources. The white (gold) symbol is the value of the global measurement over the entire galaxy. Figure from Galliano \etal\, 2008b. }
\label{peeters_fint}
\end{figure}

The relative intensities of the PAH bands vary from source to source and spatially within extended sources.  Some PAH band intensities show strong correlations with each other, revealing the underlying connection amongst the PAH bands (Fig. \ref{peeters_fint}, e.g. Hony \etal\, 2001; Vermeij \etal\, 2002; Brandl \etal\, 2006; Sellgren \etal\, 2007, Smith \etal\, 2007; Galliano \etal\, 2008b; Acke \etal\, 2010; Bernard-Salas \etal\, 2010; Boersma \etal\, 2010; Rosenberg \etal\, 2011; Peeters \etal\, 2011). Specifically, the strength of the 3.3 \mum\, feature correlates with that of the 11.2 \mum\, feature.  Similarly, the strength of the 6.2, 7.7, 8.6, 11.0, 12.7 and 16.4 \mum\, features correlate well with each other. These variations are found in the Milky Way, the Magellanic Clouds and nearby galaxies.

Few sources exhibit peculiar PAH band strength ratios. Some T-Tauri stars and HAeBe-stars show weak or no 7.7 and 8.6 \mum\,  PAH emission (Acke \& van den Ancker 2004, Geers  \etal\, 2006). Similarly, several galaxies harboring an AGN show very weak emission in the 6 to 9 \mum\, region relative to the 11.2 \mum\, band intensity (Kaneda \etal\, 2005; Smith \etal\, 2007; Bregman \etal\, 2008). Moreover, for these galaxies, the 7.7/11.2 PAH ratio depends on the hardness of the radiation field, in contrast to galaxies with \HII region or starburst-like characteristics.  

In addition to variations in the relative band intensities, the total PAH intensity is also highly variable. The intensity ratio of PAHs to Very Small Grains (VSGs, determined around 15-25 \mum) varies spatially across extended Galactic \HII regions (e.g. Lebouteiller \etal\, 2007, Churchwell \etal\, 2009). In particular, the dust continuum at 15-25 \mum\, dominates inside the \HII region while in the Photo-Dissociation Region (PDR), the PAHs emission is strongest. Similarly, the PAH/VSG ratio in galaxies depends on the hardness of the radiation field. Furthermore, the ratio of the PAH intensity to the total dust emission declines with decreasing metallicity, and hence is in addition to a decrease in total dust content in low metallicity environments (e.g. Madden \etal\, 2006; Brandl \etal\, 2006; Engelbracht \etal\, 2006; Galliano \etal\, 2008a; Gordon \etal\, 2008; Calzetti 2011 and references therein). The underlying cause of this relation is still unclear; the two main scenarios are a less efficient PAH formation process (metallicity effect) or in an increased PAH processing (modification and/or destruction of PAHs by the hard radiation field).

\subsection{Implications:}

Laboratory and theoretical spectroscopy of PAHs and related species reveal that their intrinsic spectra are set by charge, size, molecular (edge) structure and, temperature. Consequently, the observed variety in astronomical PAH spectra reflects the properties of the present PAHs. Here we give two examples.

Laboratory and theoretical studies on PAHs clearly show the large influence of the PAH charge state on their IR spectra, in particular on the band intensities (e.g. Hudgins \& Allamandola 2004 and reference therein). Neutral PAHs emit strongly at 3.3 and 11.2 \mum\, while ionized PAHs show dominant emission in the 5-10 \mum\, region. Furthermore, in the 10-15 \mum\, region, the 11.0 \mum\, PAH band is attributed to ionized PAHs and the 11.2 \mum\, band to neutral PAHs. This is consistent with the observed variations in the intensities of the main PAH bands: the degree of ionization dominates the observed range in the intensity ratios of the 6.2, 7.7 or 8.6 \mum\, PAH band and the 3.3 or 11.2 \mum\, PAH band (Joblin \etal\, 1996; Hony \etal\, 2001; Galliano \etal\, 2008b).

The observed relative intensities of the CH$_{oop}$ bands (10-15 \mum\, region) unveil the molecular edge structure of the PAHs (Hony \etal\, 2001; Bauschlicher \etal\, 2008, 2009). Specifically, the peak wavelengths of the CH$_{oop}$ bending modes depend strongly on the number of adjacent peripheral C-atoms bonded to an H-atom (Bellamy 1958; Hony \etal\, 2001). The dominance of the 11.2 \mum\, band in planetary nebulae then indicates PAHs characterized by very compact molecular structures with long smooth edges. In contrast, interstellar PAHs have more corners, either because they are on average smaller or they are more irregular larger species.
Similarly, the peak position of the CH stretching mode depends on the molecular edge structure and suggests that the edge structure of astronomical PAHs is remarkably regular and smooth (Bauschlicher \etal\, 2009). 

\begin{figure}
\hspace{-.3cm}
\includegraphics[width=7cm]{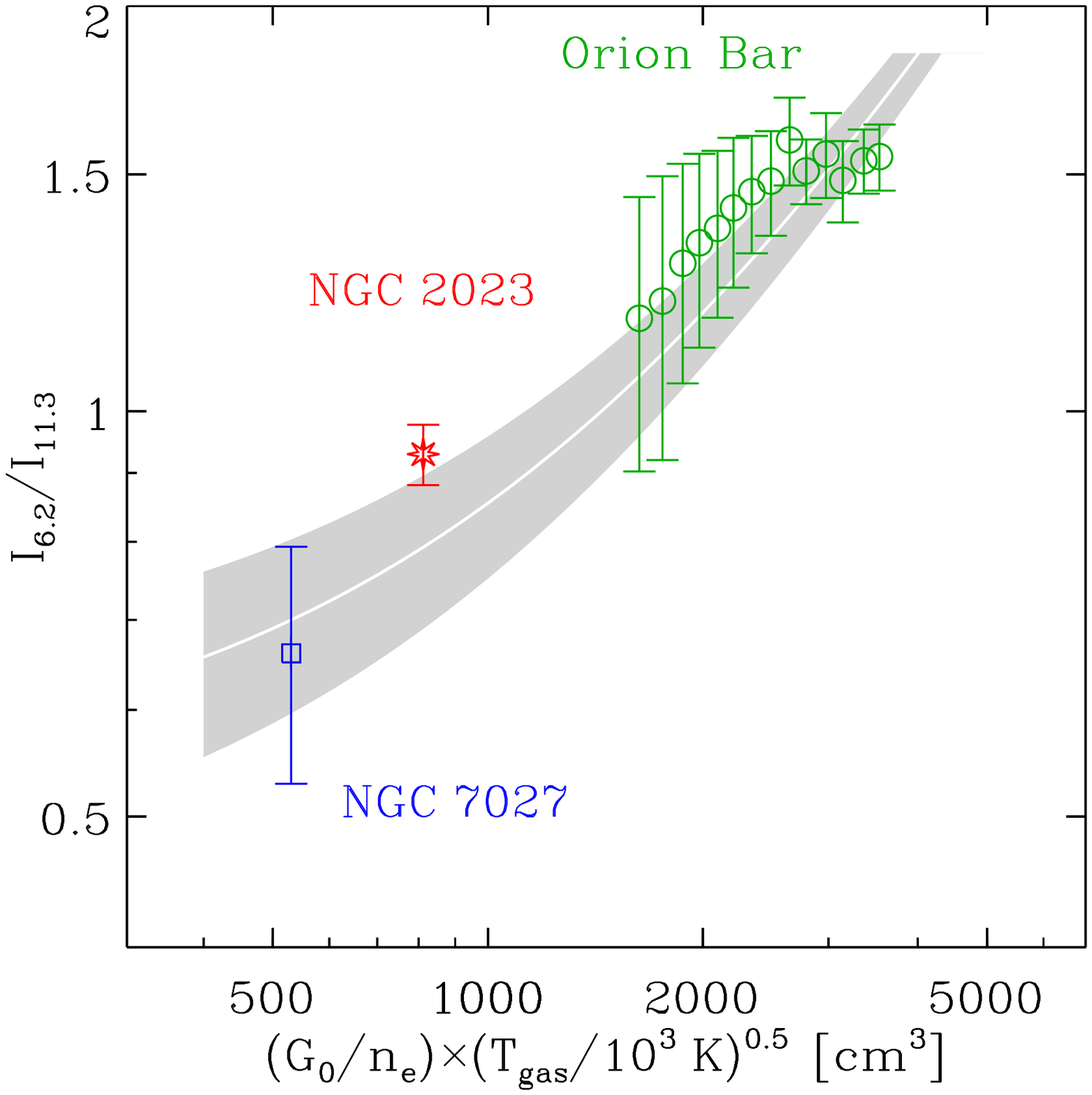}
\qquad
\hspace{-1.2cm}
\includegraphics[width=7.5cm]{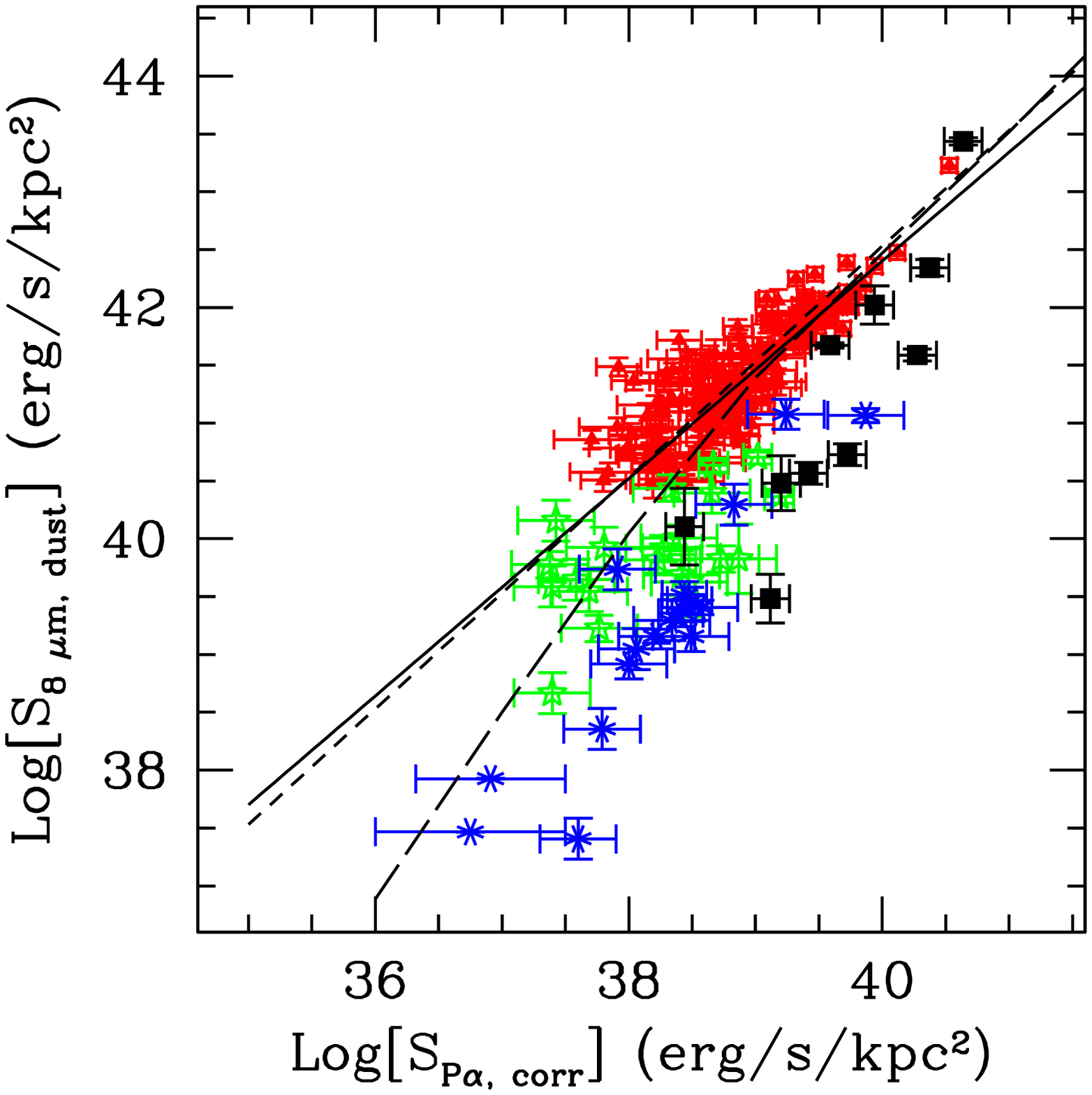}
\caption{{\bf Left:} Empirical calibration of the 6.2/11.2 band strength ratio as a function of the ionization parameter, G$_0$T$^{1/2}$/n$_e$, where G$_0$ is the UV radiation field, T the gas temperature and n$_e$ the electron density. Figure from Galliano \etal\, 2008b. {\bf Right:} The surface luminosity density at 8 \mum\, (stellar-continuum subtracted) as a function of the surface luminosity density of the extinction corrected P$\alpha$ line for low metallicity starburst galaxies (black; Engelbracht \etal\, 2005) and \HII knots. The latter are divided in 3 metallicity bins: high (red filled triangles), intermediate (green stars) and low (blue asterisks). Also shown are the best linear fit through the high metallicity data points (continuous line) and the best linear fit with unity slope (dashed line). Low metallicity regions and starburst galaxies exhibit a depressed 8 \mum\, emission due to a deficiency of PAH emission. Figure from Calzetti 2011.}
\label{peeters_tools}
\end{figure}

\subsection{PAH toolbox}

The local physical conditions determine the ionization fraction of the PAHs, which in turn specifies the (6.2, 7.7, 8.6)/11.2 PAH intensity ratio.  Hence, if the variations in the physical conditions are known (i.e. determined from other diagnostics such as e.g. PDR models), an empirical calibration of the PAH bands with the local physical conditions can be established (Fig. \ref{peeters_tools}, Galliano \etal\, 2008b, Bern\'e \etal\, 2009). Given that the PAH bands are omnipresent and easily detectable, they then become a powerful diagnostic tool for the local physical conditions. 

Given that PAHs are excited by UV radiation, the PAH emission bands are very bright in massive star-forming regions and so are a powerful tracer of star formation throughout the universe. They are utilized to determine the contribution of star formation to the total power budget of IR emission from galactic nuclei (e.g. Genzel \etal\, 1998; Lutz \etal\, 1998; Peeters \etal\, 2004b; Smith \etal\, 2007). In addition, they are extensively being explored as a quantitative probe for the star formation rates of galaxies (Fig. \ref{peeters_tools}, Calzetti 2011 and references therein). This is not that straightforward because the PAH emission depends on metallicity and, PAHs are also excited by the older stellar population, unrelated to star formation.

Finally, the presence of PAHs is used to distinguish between shocked gas and PDRs (van den Ancker \etal\, 2000) and to determine redshifts in distant galaxies (e.g. Yan \etal\, 2007).

\section{Band profiles}

\subsection{Observational facts}
The profiles of the PAH emission bands range from fairly symmetric (e.g. the 3.3 and 8.6 \mum\, bands) to highly asymmetric with a steep blue rise and a red-shaded wing (e.g., 3.4, 5.25, 6.2, 11.2 \mum; Barker \etal\, 1987; Roche \etal\, 1996; Verstraete \etal\, 2001; Pech \etal\, 2002; Boersma \etal\, 2008). In contrast, the 12.7 \mum\, band profile has blue-shaded wing with a steep red decline. Additionally, the 7.7 \mum\, complex consists out of at least 2 components: one peaking at $\sim$ 7.6 \mum\, and the second between 7.8-8 \mum\, (e.g. Bregman 1989). Further substructure may be present near 7.2-7.5 \mum\, and at 8.2 \mum\, (e.g. Moutou \etal\, 1999; Peeters \etal\, 2002).

\begin{figure}
\hspace{-1cm}
\includegraphics[width=11.cm, angle=90]{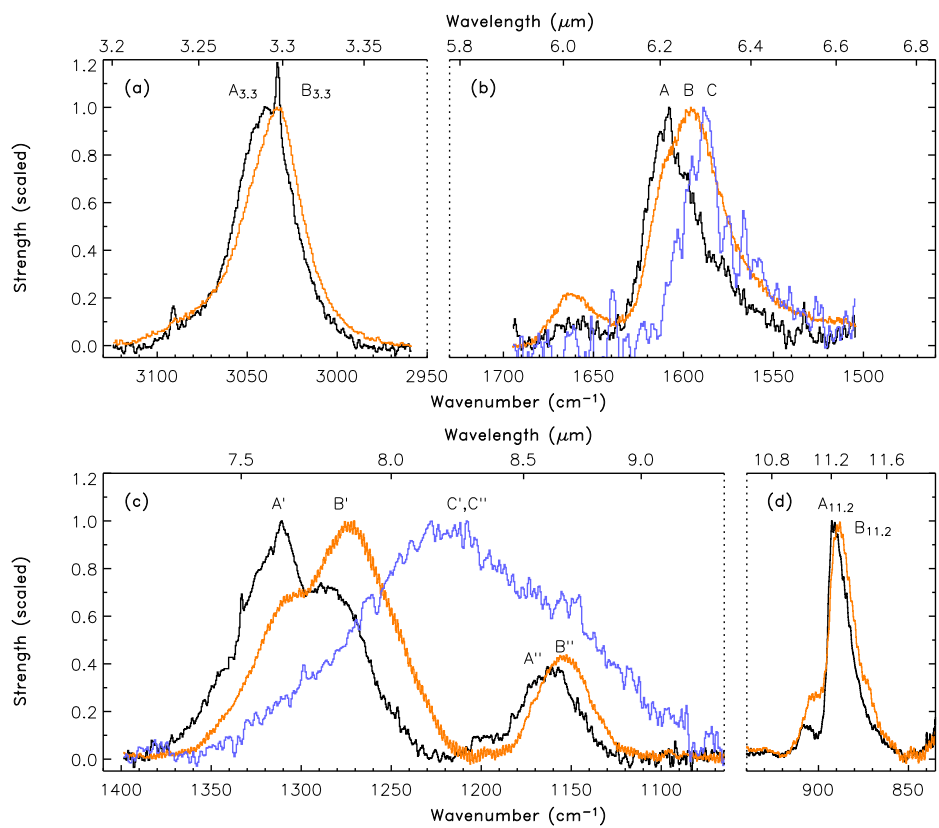}
\caption{The source to source variations in position and profile of the main PAH bands. In particular large variations are evident in the 6 to 9 \mum. Class A peaks at the shortest, 'nominal' wavelengths, class B at longer wavelengths, and class C at even longer wavelengths (Peeters \etal\, 2002; van Diedenhoven \etal\, 2004). Figure from van Diedenhoven \etal\, 2004.}
\label{peeters_profiles}
\end{figure}

The band profiles for the main PAH bands show clear variability (e.g. Bregman 1989; Cohen \etal\, 1989; Tokunaga \etal\, 1991; Peeters \etal\, 2002; van Diedenhoven \etal\, 2004). They are classified in three classes A, B and C, which is primarily based on the peak position of the bands  (see Fig. \ref{peeters_profiles}; Peeters \etal\, 2002; van Diedenhoven \etal\, 2004). For the 7.7 \mum\, complex, this results in a band profile with either a dominant 7.6 \mum\, component (class A) or a dominant component peaking between 7.8 and 8 \mum\, (class B) or a very broad band peaking at $\sim$ 8.2 \mum\, with a weak to absent 8.6 \mum\, PAH band (class C). Within this classification, class A and C show little variation while large differences in peak position and profile are present within class B. In addition, few objects have band profiles that encompasses 2 classes (AB: Van Kerckhoven 2002; Peeters \etal\, 2002, Boersma \etal\, 2008; BC: Sloan \etal\, 2007). Hence, the band profiles seem to span a continuous distribution with class A and C as the two extremes.

These band profile variations are observed in the Milky Way, and recently also in the Magellanic Clouds (Bernard-Salas \etal\, 2009, Matsuura \etal\, 2010). They are also observed within (some) extended objects (e.g. Bregman \& Temi 2005, Song \etal\, 2007). Remarkably, the 7.7 \mum\, complex varies within reflection nebulae (Bregman \& Temi 2005); however, no changes are seen within \HII regions (Galliano \etal\, 2008b). Similarly, the classes depend on object type and hence environment (Peeters \etal\, 2002, van Diedenhoven \etal\, 2004). Class A is primarily observed in \HII regions, reflection nebulae, the ISM and galaxies but includes few post-AGB stars and planetary nebulae; class B is found in post-AGB stars, most planetary nebulae and isolated Herbig AeBe stars; and class C is primarily observed towards post-AGB stars but includes few Herbig AeBe and T-Tauri stars. In other words, interstellar material exhibit class A profiles while circumstellar material (CSM) shows class B and C profiles.

\begin{figure}
\center{\includegraphics[width=6cm]{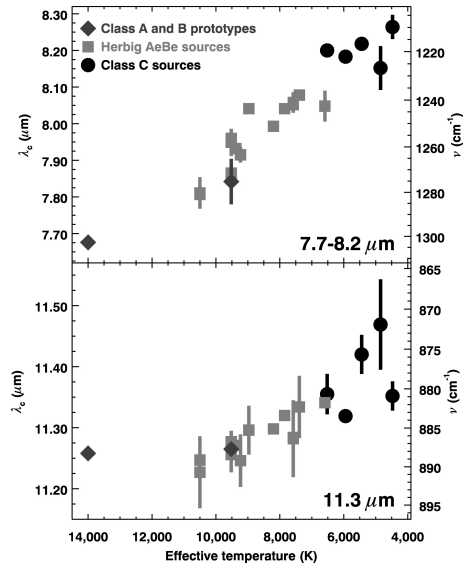}}
\caption{The central wavelengths of the 7.7--8.2 and 11.3 \mum\, PAH features plotted versus the effective temperature of the host star. Figure from Keller \etal\, 2008. }
\label{peeters_teff}
\end{figure}

 The dependence on the local environment is further established as follows. The central wavelength of the 7.7 \mum\, complex strongly correlates with the radiation field G$_0$ and the PAH ionization parameter G$_0$/n$_e$ (the ratio of the UV intensity G$_0$ to the electron density n$_e$) within reflection nebulae (Bregman \& Temi 2005). Moreover, the central wavelength of the main PAH bands shows a remarkable anti-correlation with the effective temperature of the exciting star for a sample of post-AGB stars and isolated HAeBe stars, all of class B or C (Fig. \ref{peeters_teff}; Sloan \etal\, 2007; Keller \etal\, 2008; Acke \etal\, 2010). This does not seem to hold in general. First, some class A objects, such as reflection nebulae for example, have a central star with low effective temperatures. Secondly, some planetary nebulae have high effective temperatures and class B profiles (e.g. NGC7027). Third, the PAH bands of Herbig AeBe stars with the same effective temperature can belong to different classes (Van Kerckhoven 2002; Boersma \etal\, 2008). This suggests that other parameters such as e.g. history, environment (ISM vs CSM), spatial structure (disk, collapsing cloud; Boersma \etal\, 2008) may play a role.

\subsection{Implications:}

These band profile variations primarily originates in a (chemical) modification of the PAHs. However, the specifics of this modification is still under debate. Recent studies focus on the CC stretching bands because i) the position of the class A 6.2 \mum\, PAH band cannot be reproduced by the strongest pure CC stretching mode of pure PAHs (Peeters \etal\, 2002; Hudgins \etal\, 2005) using the current available laboratory and theoretical data, and ii) the 7.7 \mum\, complex exhibit the most pronounced variability.  Various PAH-related species and processes have been proposed:
\begin{itemize}
\item {\it Hetero-atom substituted PAHs} (e.g. Peeters \etal\, 2002; Hudgins \etal\, 2005; Bauschlicher \etal\, 2009). In particular, the substitution of a C atom by a N atom in a PAH, i.e. PANHs. 
\item  {\it PAH-metal complexes} (e.g. Hudgins \etal\, 2005; Bauschlicher \& Ricca 2009; Simon \& Joblin 2010; Joalland \etal\, 2009). These are species in which a metal atom is located either below or above the carbon skeleton. 
\item {\it PAH clusters} such as dimers and trimers (e.g. Rapacioli \etal\, 2005; Simon \& Joblin 2009). 
\item  {\it Variations in the size distribution of the PAH family} (Bauschlicher \etal\, 2008, 2009; Cami 2011; Cami \etal\, 2011). Small PAHs ($N_C$ $\le$ 48) emit at $\sim$7.6 \mum\, while large PAHs (54 $\le$ $N_C$ $\le$ 130) emit at $\sim$7.8 \mum. The broad class C component, peaking at $\sim$8.2 \mum, can be modelled with a collection of small PAHs (Sect. \ref{peeters_decomposition}). Hence, the PAH classes may then reflect different size distributions of the emitting PAHs. 
\item  {\it Varying importance of aliphatics vs. aromatics} (e.g. Sloan \etal\, 2007; Boersma \etal\, 2008; Keller \etal\, 2008; Pino \etal\, 2008; Acke \etal\, 2010).  Aliphatic carbonaceous species are proposed for the carrier of the class C 7.7 \mum\, complex. UV processing of these species destroys the aliphatic bonds and hence increases their aromaticity. Different degrees of UV processing (e.g. due to an increase in effective temperature) is then invoked to explain the different classes. Since UV processing may also influence the PAH size distribution, both mechanisms may occur together. While increased UV processing can explain the evolutionary scenario from post-AGB stars to the ISM, the reversed process from aromatic material to more fragile aliphatic material is hard to image. The latter occurs when going from the ISM to proto-planetary environments. Hence, an active chemical equilibrium between aromatic and aliphatic species is proposed for all environments through hydrogenation, carbon reactions building (aliphatic) hydrocarbons and UV processing (Boersma \etal\, 2008).

\end{itemize}
\vspace{-.4cm}

\section{Spectral decomposition}
\label{peeters_decomposition}

\begin{figure}
\includegraphics[width=12cm]{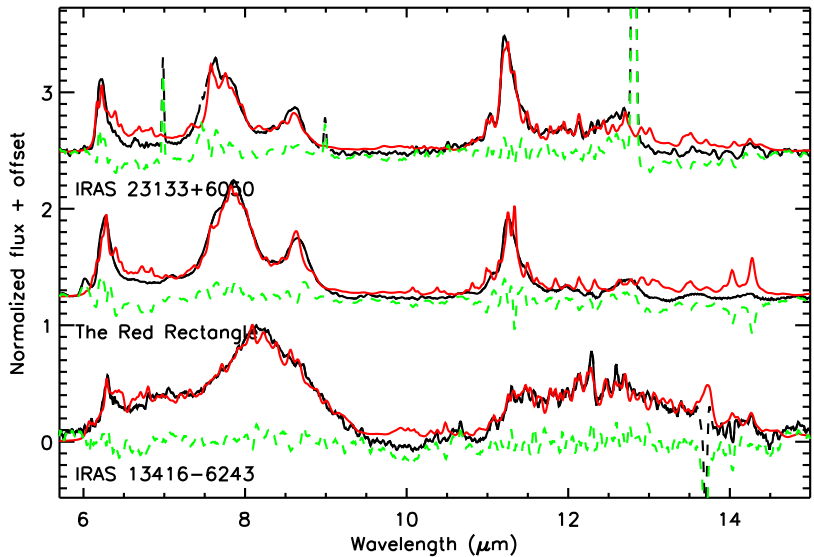}
\caption{The astronomical PAH spectra (black) and the best fit
PAH model spectra (ligther curves). Residuals are shown as dashed lines. The different classes are represented by the \HII region IRAS 23133+6050 (class A, top), the Red
Rectangle (class B, middle), and the proto-planetary nebula IRAS 13416-6243
(class C, bottom). Figure from Cami 2011.}
\label{peeters_fit}
\end{figure}

Over the years, a large amount of spectra from laboratory experiments and computational modeling of PAHs are obtained. Many of these are publicly available at the NASA Ames PAH spectral database (Bauschlicher \etal\, 2010) and the Cagliari/Toulouse database\footnote{At http://www.astrochem.org/pahdb/ and http://astrochemistry.ca.astro.it/database respectively.}. These databases can now be used to model the astronomical PAH spectra. Amazingly good fits are obtained to all three classes A, B and C with the full theoretically calculated NASA Ames PAH spectral database (Fig. \ref{peeters_fit}; Cami 2011, Cami \etal\, 2011). Although these fits are not unique and influenced by the bias present in the database, they can be broken down according to size, charge and composition and thus unveil the underlying origin of the spectral variations. For example, they show that the 6.2 \mum\, band requires a significant contribution of nitrogen-substituted PAHs; the 7.7 \mum\, band profile reflects the PAH size distribution; the 11.2 \mum\, band is due to large, neutral and pure PAHs; and the 11.0 \mum\, band is due to large PAH cations.

\begin{figure}
\center{\includegraphics[width=8cm]{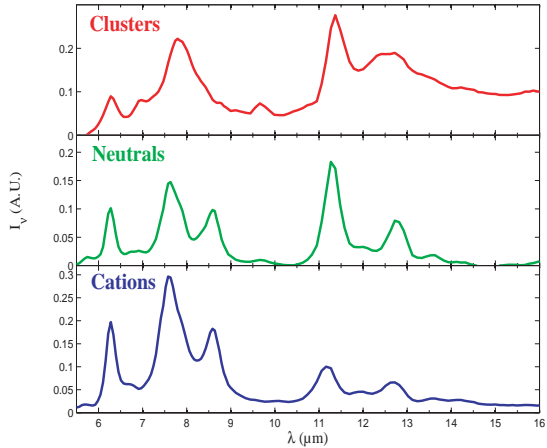}}
\caption{The three components in the singular value decomposition analysis of NGC 7023: PAH$^+$ (bottom), PAH$^0$ (middle) and VSG or PAH clusters (top). The components are normalized by their integrated intensity. Figure adapted from Rapacioli \etal\, 2005.}
\label{peeters_bss}
\end{figure}

The mathematical method Blind Signal Separation is applied to spectral maps of PAH spectra in order to decompose the PAH spectra as a linear combination of a basis set of spatially different components (Boissel \etal\, 2001; Rapacioli \etal\, 2005; Bern\'{e} \etal\, 2007, 2009).  These studies found a basis set of three mathematically distinct components, referred to as PAH$^0$, PAH$^+$, and VSG\footnote{This VSG component is {\it not} the same as the VSGs discussed in Sect. \ref{peeters_intensities}.}, which have different spectral characteristics (Fig. \ref{peeters_bss}). Although this is a pure mathematical decomposition, the PAH$^0$ and PAH$^+$ components are attributed to respectively ÔneutralÕ and ÔionizedÕ PAHs and the VSG component to PAH clusters. Nevertheless, fitting of the PAH$^0$ and PAH$^+$ components with the NASA Ames PAH database confirms the attribution to neutral and ionized PAHs (Rosenberg \etal\, 2011). The PAH$^0$ and PAH$^+$ components are found closer to the edge of the PDR and the VSG component is located deeper into the PDR suggesting the destruction of the VSGs or PAH clusters by UV photons and subsequent formation of PAHs. These PAH clusters may be reformed in the denser and more shielded environments of molecular clouds. 
This method has only been applied to class A sources. Hence, in order to fit PAH spectra of class B and C, this basis set is expanded to include ÔartificialÕ templates (i.e. they do not result from a mathematical decomposition; Joblin \etal\, 2008). These four templates exhibit a very broad band at 8.2, 8.3 and 12.3 \mum\, (thus mimicking class C) and a third PAH template, PAH$^x$, which has peak positions consistent with class B.

\section{The PAH hypothesis after 25 years}

Although not a single PAH molecule has been identified to date\footnote{Mainly because the mid-IR vibrational modes of PAHs show only a weak dependence on size and structure.}, the PAH hypothesis has not been refuted. One of its strongest criticisms has been the lack of a good match between a combination of experimental and/or theoretical PAH spectra and the astronomical PAH spectra. This criticism has been addressed and is now no longer in play (Sect. \ref{peeters_decomposition}). However, weaknesses in the original ÒPAH hypothesisÓ have been revealed resulting in a refinement of the ÒPAH hypothesisÓ. Specifically, it now includes ÔpureÕ PAHs as well as PAH-related species such as PAHs containing impurities and (aliphatic) side-groups, hydrogenated PAHs, PAHs clusters, etc. Hence, a strict chemical definition of ÔPAHsÕ is no longer in use and it is arguable whether these species can really be dubbed PAHs. Nevertheless, this refined ÒPAH hypothesisÓ is alive and kicking.

To date, the wealth of astronomical PAH spectra and its large variability prompt more questions than can be answered with the current laboratory and theoretical data. It is clear that significant progress in our understanding of the astronomical PAH spectra can only be made by a joint effort of the observational, experimental and theoretical tools. 
\newpage
\acknowledgement
We thank O. Bern\'{e}, J. Cami, D. Calzetti, F. Galliano, S. Hony, L. Keller and M. Rapacioli for providing figures for this article. Figures are reproduced by permission of the AAS, ESO and, EDP sciences. 

\begin{discussion}

\discuss{Krelowski}{You have not mentioned that PAHs should have features in the visible and that the latter remain undetected.}
\discuss{Peeters}{That's correct. There is a talk later today by Nick Cox that will address this issue.}

\discuss{Geballe}{You showed very accurate fits of an ensemble of PAHs to the 10 \mum\, PAH spectrum. What is the situation regarding simultaneous fitting of the 3.3 and 3.4 \mum\, features along with the 10 \mum\, bands?}
\discuss{Peeters}{The intensity of the 3 \mum\, bands in the theoretical spectra is off by a factor of $\sim$2. Hence, they currently cannot be fitted together with the bands at longer wavelengths.}

\discuss{Menten}{For fitting the observed spectra with templates from the PAH database, do you have templates that were produced for different temperatures and do you use a single temperature for all the template spectra of different species?}
\discuss{Peeters}{No, we don't have temperature information for most PAHs in the database.}

\end{discussion}

\end{document}